\documentclass[pre,twocolumn,superscriptaddress,floatfix]{revtex4}
\usepackage{graphicx}

\newcommand{\be}{\begin{equation}}
\newcommand{\ee}{  \end{equation}}
\newcommand{\ba}{\begin{eqnarray}}
\newcommand{\ea}{  \end{eqnarray}}

\begin{document}

\title{Chaos in Fermionic Many--Body Systems and the Metal--Insulator
Transition}

\author{T. Papenbrock}
\affiliation{Department of Physics and Astronomy, University of Tennessee,
Knoxville, TN 37996, USA}
\affiliation{Physics Division, Oak Ridge National Laboratory, Oak Ridge,
TN 37831, USA}

\author{Z. Pluha\v r}
\affiliation{Institute of Particle and Nuclear Physics, Faculty of
Mathematics and Physics, Charles University, 18000 Praha 8, Czech
Republic}

\author{J. Tithof}
\affiliation{Department of Physics and Astronomy, University of Tennessee,
Knoxville, TN 37996, USA}

\author{H. A. Weidenm{\"u}ller}
\affiliation{Max-Planck-Institut f{\"u}r Kernphysik, D-69029 Heidelberg,
Germany}

\begin{abstract}
We show that finite Fermi systems governed by a mean field and a
few--body interaction generically possess spectral fluctuations of the
Wigner--Dyson type and are, thus, chaotic. Our argument is based on an
analogy to the metal--insulator transition. We construct a sparse
random--matrix ensemble ScE that mimics that transition. Our claim
then follows from the fact that the generic random--matrix ensemble
modeling a fermionic interacting many--body system is much less sparse
than ScE.
\end{abstract}

\maketitle

\section{Introduction}

Finite fermionic many--body systems (atoms, molecules, nuclei) often
display spectral fluctuation properties that agree with predictions of
random--matrix theory (RMT), more precisely, with those of the
Gaussian Orthogonal Ensemble (GOE)~\cite{Guh98}. This fact is commonly
taken as evidence for chaotic motion~\cite{Boh84,Heu07}. In RMT, all
pairs of states are coupled by independent matrix elements. In a
many--body system that situation arises only in the presence of
many--body interactions. But atoms, molecules and nuclei are governed
by a mean field with residual interactions that are predominantly of
two--body nature. Therefore, an important question is: Does a
two--body interaction (or, more generally, a $k$--body interaction
with $k \geq 2$ integer but smaller than the number $m$ of fermions)
generically give rise to chaotic motion? The question has received
much attention (see the review~\cite{Ben03}). In condensed--matter
physics it has recently been addressed as the problem of many--body
localization~\cite{Bas08}. 

While for small values of $m$ the question can be answered by matrix
diagonalization, a general answer is difficult to obtain both
analytically and numerically because the Hamiltonian matrices become
ever more sparse with increasing $m$: In every row and column the
ratio of the number of non--zero matrix elements to the total number
of matrix elements tends asymptotically ($m \to \infty$) to zero.
Therefore, analytical arguments like the ones in Ref.~\cite{Fyo91}
(where that ratio is taken to be asymptotically finite) do not
apply. And for the physically interesting cases $k \geq 2$ matrices
with dimensions that are numerically practicable are very far from the
sparse limit.

The most determined effort so far to overcome these difficulties was
made in Ref.~\cite{Jac97}. The authors considered a fermionic
many--body system governed by random one-- and two--body
interactions. For fixed parameter values of the one--body part, they
determined the critical strength of the two--body interaction where
the crossover from Poisson statistics to Wigner--Dyson statistics
takes place. Their arguments are based on a combination of a
perturbative approach and numerical results. Their result implies that
a system governed by a pure two--body interaction is chaotic. Needless
to say, the matrices studied numerically were far from the sparse
limit.

In this paper we aim at a further clarification of the issue. We
consider the matrix representation of a random Hamiltonian of the
$k$--body type. We are interested in the way in which properties of
the resulting ensemble of random matrices affect both, the shape of
the average spectrum and the spectral fluctuation properties. Our
approach differs from previous ones in that we are guided by an
analogy to the metal--insulator transition (MIT). In disordered
metals, the degree of disorder determines whether a system is an
insulator (with Poissonian level statistics) or a metal (with
Wigner--Dyson level statistics). As disorder increases, a transition
from the metallic to the insulating regime takes place~\cite{And58}.
At the transition point, the spectral fluctuation properties are
governed by a ``critical statistic'' which usually is characterized by
three measures~\cite{Eve08}: (i) The distribution of spacings $s$ of
nearest eigenvalues is linear in $s$ for small $s$ and falls off
exponentially for large $s$; (ii) the variance $\Sigma^{(2)}(L)$ of
the number of eigenvalues in an interval of length $L$ is
approximately logarithmic for small $L$ and linear for large $L$;
(iii) the eigenfunctions display fractional statistics, and the
distribution of the inverse participation ratios (sum of the fourth
power of the expansion coefficients of the eigenfunctions in an
arbitrary basis, see Eq.~(\ref{ipr}) below) scales with $N$ like an
inverse fractional power of $N$. Properties (i) and (ii) obviously
interpolate between Wigner--Dyson statistics for short spectral
distances and Poisson statistics for larger spectral distances.

Within the framework of random--matrix theory, an ensemble has been
constructed~\cite{Mir96} that simulates the critical statistic at the
MIT. That ensemble is a special case of a power--law random band
matrix (PLRBM). In such random matrix ensembles, the variances of the
non--diagonal matrix elements $H_{\mu \nu}$ fall off with some power
$2 a$ of the distance $|\mu - \nu|$ from the main diagonal. For $a >
1$ the spectral statistics of the PLRBM is Poissonian, for $a < 1$ it
is of Wigner--Dyson type, and for $a = 1$ it is critical~\cite{Mir96}.
We note that in contrast to Hamiltonian matrices for interacting Fermi
systems the PLRBM is not sparse.

Motivated by the analogy to the MIT and by the PLRBM in
Ref.~\cite{Mir96}, we construct a random--matrix ensemble (the
``scaffolding ensemble'' [ScE]) with the following properties. (i) ScE
is more sparse than the Hamiltonian matrix of a fermionic many--body
system with $k$--body interactions. (ii) The spectral fluctuation
properties of ScE are those of the critical ensemble. This then
suggests (and we show) that for all $k \geq 2$, the Hamiltonian matrix
of the fermionic problem lies on the metallic side of the MIT and is,
therefore, chaotic. The case $k = 1$ is special and supplies
additional arguments that support our reasoning.

Not surprisingly we cannot offer strict analytical proofs for some of
these statements. Our arguments are based on a combination of
analytical arguments, numerical evidence, and the application of a
criterion due to Levitov~\cite{Lev90} that is introduced below.

\section{Embedded Ensemble (EGOE($k$))}
\label{EGOE}

Specifically we investigate a paradigmatic random--matrix model that
simulates a fermionic many--body system: The Embedded Gaussian
Orthogonal Ensemble (EGOE)~\cite{Mon75}. We consider $m$ spinless
Fermions in $l > m$ degenerate single--particle states labeled $j =
1, \ldots, l$ with associated creation and annihilation operators
$a^{\dag}_j$ and $a^{}_j$, respectively. The states carry no further
quantum numbers.  With $1 \leq k \leq m$ the $k$--body Hamiltonian is
\be
{\cal H}^{({k})} = \frac{1}{k!^2} \sum_{j_1 \ldots j_k; j'_1 \ldots
j'_k} v^{j_1 \ldots j_k}_{j'_1 \ldots j'_k} a^\dag_{j_1} \ldots
a^\dag_{j_k} a^{}_{j'_k} \ldots a^{}_{j'_1} \ .
\label{p4}
\ee
The matrix $v$ is real--symmetric. The matrix elements are
antisymmetric under the exchange of any pair of primed or unprimed
indices and are uncorrelated random variables with a Gaussian
probability distribution with zero mean value and a common second
moment which without loss of generality is taken to be unity. The
matrix representation of ${\cal H}^{(k)}$ in the space of $m$--body
Slater determinants labeled $\mu$ or $\nu$ with $m$--body matrix
elements $\langle \nu| {\cal H}^{(k)} | \mu \rangle = H_{\nu \mu}$
defines an ensemble of real random matrices of dimension $N = {l
  \choose m}$ referred to as EGOE($k$). We study the spectral
properties of EGOE($k$) in the limit $N \gg 1$. For fixed $k$ that
limit is reached by letting $l, m \to \infty$. Without distinction we
consider both the dilute limit ($m / l \to 0$) and the dense limit ($m
/ l \to$ constant $\neq 0$). In Dyson's classification, ${\cal
  H}^{(k)}$ has orthogonal symmetry. Our arguments apply likewise to
the cases of unitary and symplectic symmetry. As mentioned above, for
$k < m$ EGOE($k$) has withstood all attempts at a direct analytical
treatment~\cite{Ben03,MS02}.

\section{Scaffolding Ensemble (ScE)}
\label{ScE}

We first describe the construction of the scaffolding ensemble. A
criterion due to Levitov then suggests and numerical simulations
confirm that ScE possesses the critical behavior characteristic of the
MIT.

We define ScE with the help of a real and symmetric scaffolding matrix
$A^{(n)}$ that has dimension $N = 2^n$ with $n$ positive integer. We
recall the definition of the auxiliary diagonal for matrices of
dimension $N$: The auxiliary diagonal has matrix elements with indices
$(\mu, N + 1 - \mu)$ and $\mu = 1, \ldots, N$. We construct the matrix
$A^{(n)}$ by induction: For $n = 1$, $A^{(1)}$ has dimension $2$, zero
diagonal elements and unit entries in the auxiliary diagonal. Given
$A^{(n - 1)}$, the two diagonal blocks (dimension $2^{(n - 1)}$) of
the matrix $A^{(n)}$ are each occupied by $A^{(n - 1)}$. The elements
in the two off--diagonal blocks are all zero except for the auxiliary
diagonal for which all elements have the value unity. We display the
scaffolding matrix $A^{(n)}$ for $n = 3$ as an example,
\be
A^{(3)} = \left( \matrix{ 0 & 1 & 0 & 1 & 0 & 0 & 0 & 1 \cr
                          1 & 0 & 1 & 0 & 0 & 0 & 1 & 0 \cr
                          0 & 1 & 0 & 1 & 0 & 1 & 0 & 0 \cr
                          1 & 0 & 1 & 0 & 1 & 0 & 0 & 0 \cr
                          0 & 0 & 0 & 1 & 0 & 1 & 0 & 1 \cr
                          0 & 0 & 1 & 0 & 1 & 0 & 1 & 0 \cr
                          0 & 1 & 0 & 0 & 0 & 1 & 0 & 1 \cr
                          1 & 0 & 0 & 0 & 1 & 0 & 1 & 0 \cr} \right)
\ .
\label{p5}
\ee
By construction, the matrices $A^{(n)}$ have the important property
$\sum_{\nu = 1}^N A^{(n)}_{\mu \nu} = \sum_{\mu = 1}^N A^{(n)}_{\mu
\nu} = n$ for all $\mu = 1, \ldots, N$. Thus, in every row and column
of $A^{(n)}$, the number of non--zero non--diagonal elements is $n =
\ln N/ \ln 2$. Hence, with increasing distance $|\mu - \nu|$ from the
main diagonal, the average density of non--diagonal elements of
$A^{(n)}$ falls off like $1 / |\mu - \nu|$. The matrices $A^{(n)}$
share that important property with the power--law random band
matrices~\cite{Mir96} that simulate the MIT. Moreover, the matrices
$A^{(n)}$ bear some similarity to the ``ultrametric'' matrices studied
recently~\cite{Fyo10}.

With the help of $A^{(n)}$, ScE is defined for every $n$ as an
ensemble $H^{(n)}$ of random matrices. The non--zero elements of
$H^{(n)}$ reside on the unit elements of the matrix $A^{(n)}$ (hence
the name ``scaffolding matrix'' for $A^{(n)}$) and on the main
diagonal. Except for the symmetry condition $H^{(n)}_{\mu \nu} =
H^{(n)}_{\nu \mu}$ the matrix elements are uncorrelated real random
variables with a Gaussian distribution and zero mean values. With
$\alpha$ positive and
\be
B^{(n)}_{\mu \nu} = \alpha \delta_{\mu \nu} + A^{(n)}_{\mu \nu} \ ,
\label{p6}
\ee
the variances are given by
\be
(1 + \delta_{\mu \nu}) \langle H^{(n)}_{\mu \nu} H^{(n)}_{\rho \sigma}
\rangle = (\delta_{\mu \sigma} \delta_{\nu \rho} + \delta_{\mu \rho}
\delta_{\nu \sigma}) B^{(n)}_{\mu \nu} \ .
\label{p1}
\ee
In particular, the variance of every diagonal element is equal to
$\alpha$.

We address the spectral properties of ScE. With $E$ the energy and
$G(E) = 1 / (E^+ - H^{(n)})$ the retarded Green's function, the
average level density is $\rho(E) = - (1 / \pi) \Im \langle G(E)
\rangle$. Here and in what follows, angular brackets denote the
ensemble average. To calculate $\langle G(E) \rangle$, we expand
$G(E)$ in powers of $H^{(n)}$ and use Wick contraction in each term of
the sum. Following Ref.~\cite{Mon75} we denote Wick--contracted pairs
of matrix elements by the same letter and distinguish nested and
cross--linked contributions. Among the sixth--order contributions, for
instance, $ABCCBA$ and $ABBACC$ are nested while $ABCABC$ and $ABABCC$
are cross--linked. For $n \gg \alpha$, only nested contributions
contribute to $\langle G(E) \rangle$. That rule (for the case of the
GOE demonstrated in Ref.~\cite{Mon75}) can be inferred by comparing
the values of low--order terms like $ABBA$, $ABBACC$, and $ABCABC$.
Resummation of the nested contributions gives the Pastur equation
$\langle G(E) \rangle$ $= (1 / E) + (1 / E) \langle H^{(n)} \langle
G(E) \rangle H^{(n)} \rangle \langle G(E) \rangle$. We use
Eq.~(\ref{p1}) and find
\be
[E - \sum_\rho B^{(n)}_{\mu \rho} \langle G(E) \rangle_{\rho \rho}]
\langle G(E) \rangle_{\mu \nu} = \delta_{\mu \nu} \ .
\label{p2}
\ee
To solve Eq.~(\ref{p2}) we observe that $\langle G(E) \rangle$ is
expected to be an analytic function in $E$ with a finite number of
branch points but without singularity at $E = \infty$. Therefore, we
expand $\langle G(E) \rangle$ for $|E| \gg 1$ in a Laurent series,
$\langle G(E) \rangle_{\mu \nu} = \sum_{p = 0}^\infty E^{j - p}
g^{(p)}_{\mu \nu}$. Inserting that into Eq.~(\ref{p2}) and comparing
powers of $E$ we find that non--vanishing solutions exist only for $j
= \pm 1$. For both solutions we find that the coefficients
$g^{(p)}_{\mu \nu}$ are proportional to the unit matrix for all
$p$. That conclusion hinges in an essential way on the fact that
$\sum_\nu A^{(n)}_{\mu \nu} = n$ for all $\mu$ so that $\sum_{\rho}
B^{(n)}_{\mu \rho} = \alpha + n$ for all $\mu$, see Eq.~(\ref{p6}).
Thus $\langle G(E) \rangle_{\mu \nu} = \delta_{\mu \nu} g(E)$. To
determine $g(E)$ we use Eq.~(\ref{p2}) and find with $\lambda^{\rm sc}
= (n + \alpha)^{1/2}$ that $\lambda^{\rm sc} g(E) = (E / (2
\lambda^{\rm sc})) \pm i \sqrt{ 1 - (E / (2 \lambda^{\rm
sc}))^2}$. The two solutions with $j = \pm 1$ correspond to the two
signs in front of the square root. We conclude that the average
spectrum has the shape of Wigner's semicircle, half the GOE radius
$\lambda^{\rm GOE} \propto \sqrt{N}$ being replaced by $\lambda^{\rm
sc} = \sqrt{n + \alpha} \approx \sqrt{n}$.

We have not been able to establish the spectral fluctuation properties
of ScE (and of EGOE($k$)) analytically. For the EGOE($k$) we instead
use a criterion established in Ref.~\cite{Lev90}. To test the
applicability of that criterion to sparse random matrices, we apply it
to ScE. Levitov investigated a class of random matrices $H^{({\rm
L})}_{\mu \nu}$ for which the variances $\langle |H^{({\rm L})}_{\mu
\mu}|^2 \rangle$ of the diagonal elements are much larger than the
variances $\langle |H^{({\rm L})}_{\mu \nu}|^2 \rangle$ ($\mu \neq
\nu$) of the non--diagonal elements. All of the latter differ from
zero. Levitov considered the sum $S = \sum_\nu (1 - \delta_{\mu \nu})
\langle |H^{(L)}_{\mu \nu}|^2 \rangle^{1/2} / \langle |H^{(L)}_{\mu
\mu}|^2 \rangle^{1/2}$ in the limit of large matrix dimension
$N$. Using renormalization--group arguments, he distinguished three
cases: (i) $S$ has a finite limit. Then, the spectral statistics of
the ensemble is Poissonian.  (ii) $S$ diverges more strongly than $\ln
N$. Then, the spectral statistics is of Wigner--Dyson type. (iii) $S$
diverges like $\ln N$. Then the spectral statistics is that of the
critical ensemble at the metal--insulator transition.

ScE and EGOE($k$) are sparse random--matrix ensembles, and the
renormalization--group argument used to establish Levitov's criterion
is not readily applicable. Thus it is not clear whether the criterion
applies. If it does, ScE must for $\alpha^{1/2} \gg 1$ possess the
spectral statistics of the critical ensemble since $\sum_\nu
A^{(n)}_{\mu \nu} = n = \ln N/\ln 2$. In a finite system the
interesting regime is thus $n\gg\sqrt{\alpha}\gg 1$. Fortunately, this
regime is just within reach of present--day numerical simulations.

We construct a number of realizations of the ScE and compute spectra
via matrix diagonalization. We are limited to $n \le 13$, and we
consider the ScE for $n = 10, \ldots, 13$. Our random--matrix
ensembles consist of about 50 realizations for $n=13$ and of up to
30,000 realizations for $n = 10$. The wave--function statistics
requires the largest ensembles. For both spectral statistics and
wave--function statistics, we employ only the levels that are in the
central 20 \% of the spectral densities. In this window, the average
level density is maximal and nearly constant, and finite--size effects
can be neglected.  For the long--range $\Sigma^{(2)}$ statistic, we
verified that the results are insensitive to the degree of the
polynomial fit of the average level density.

\begin{figure}[htbp]
\includegraphics[width=0.48\textwidth,clip=]{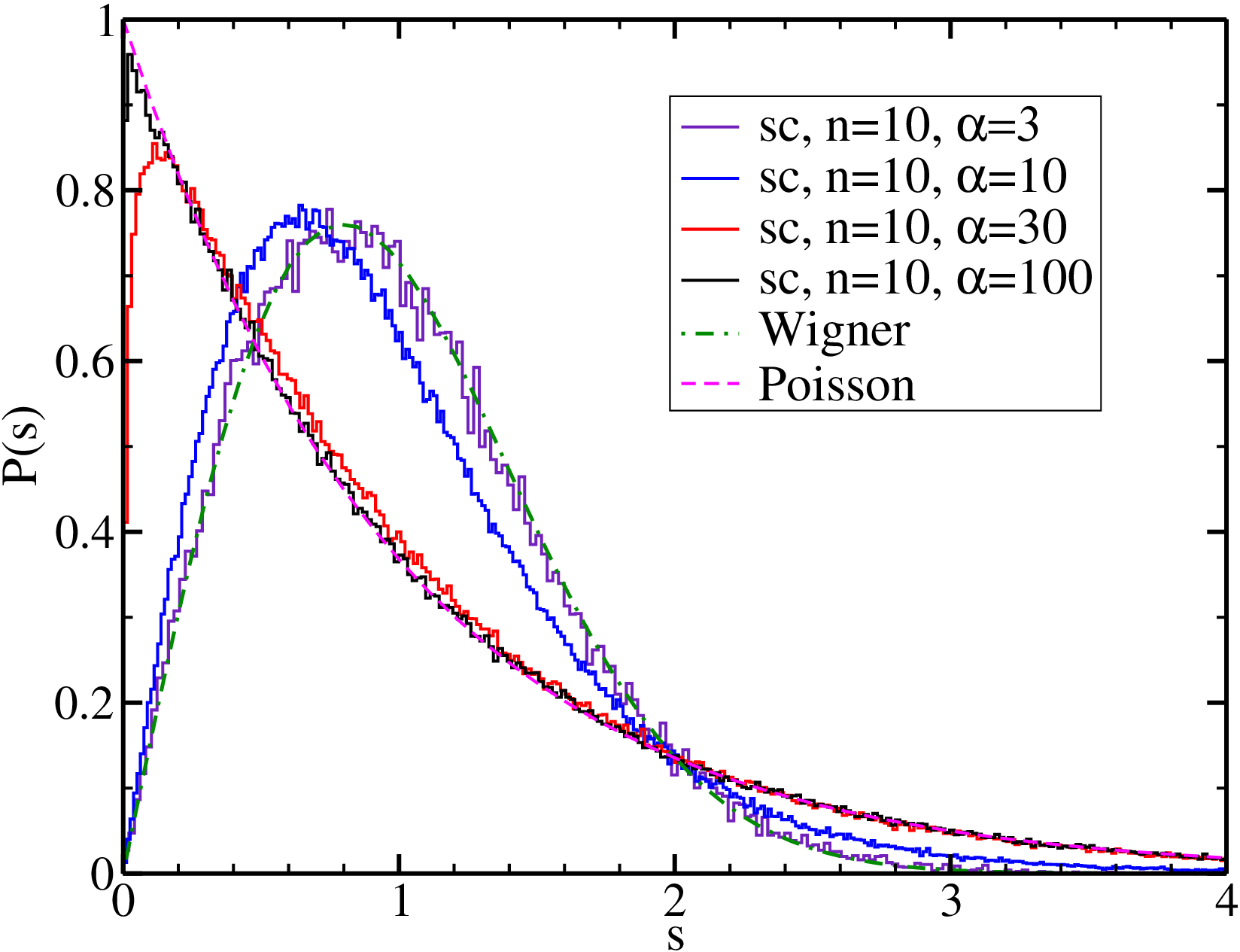}
\caption{(Color online) Nearest--neighbor spacing distribution $P(s)$
  of the scaffolding ensemble for $n = 10$ and $\alpha=3, 10, 30, 100$
  (histograms from right to left) versus $s$, the actual level spacing in units of the mean level
  spacing, compared to the Poisson distribution and Wigner's surmise
  (i.e., the GOE).}
\label{fig:n10alphavar}
\end{figure}

We first vary $\alpha$ at fixed $n=10$ and study the nearest--neighbor
spacing distribution. Figure~\ref{fig:n10alphavar} shows the
transition from a delocalized Wigner--Dyson regime to a localized
Poisson regime as $\alpha$ is increased from $\alpha = 3$ to $\alpha =
100$. (The ensembles consist of 300 realizations for $\alpha = 3$ and
of 3000 realizations for the other values of $\alpha$.)
Figure~\ref{fig:n10alphavar} suggests that we are in the critical
regime somewhere around $\alpha \approx 10$.

\begin{figure}[htbp]
\includegraphics[width=0.48\textwidth,clip=]{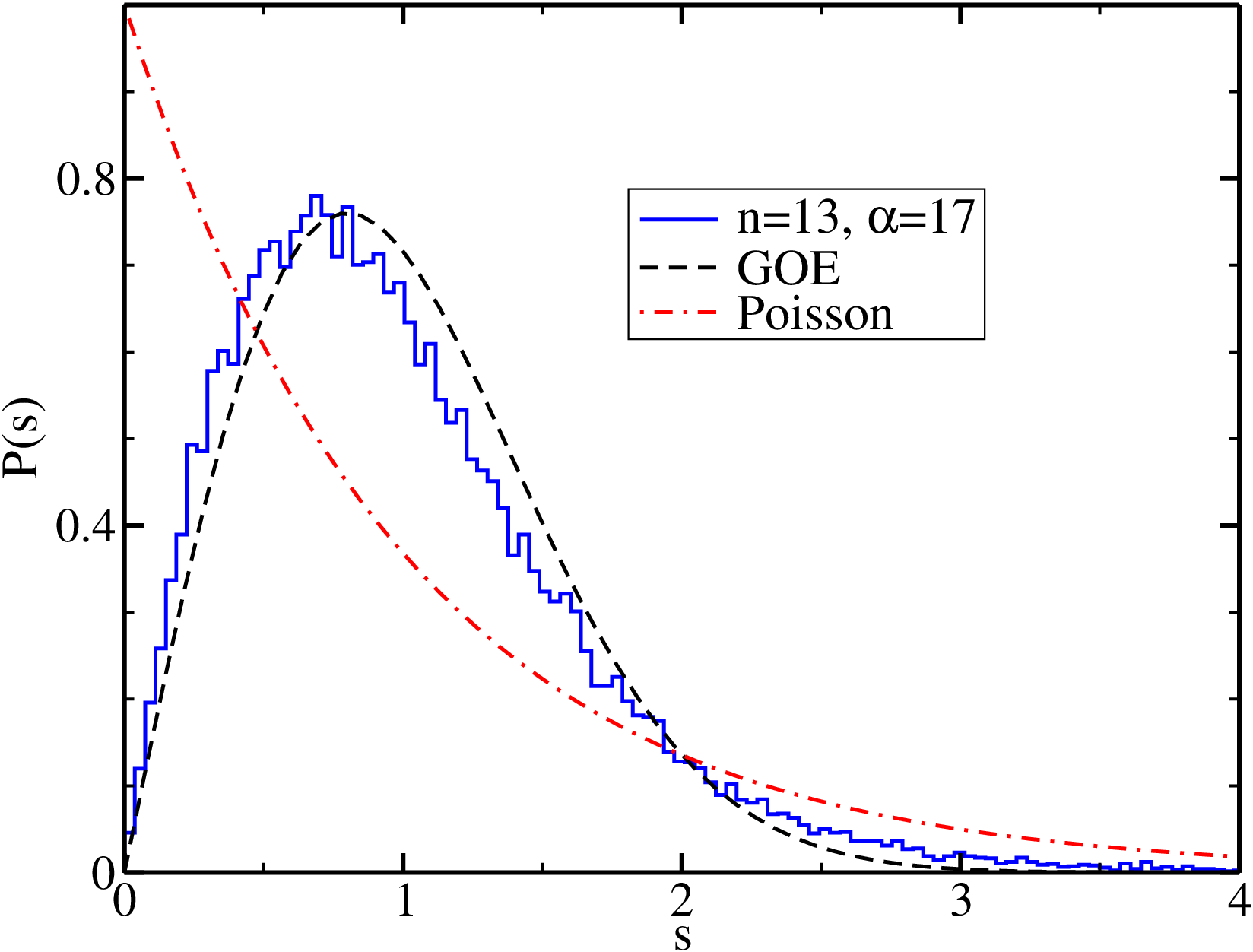}
\caption{(Color online) Nearest--neighbor spacing distribution $P(s)$
  as in Fig.~\ref{fig:n10alphavar} of the scaffolding ensemble for $n
  = 13$ and $\alpha = 17$ compared to the Poisson distribution and
  Wigner's surmise for the GOE.}
\label{fig:nns_n13_al17}
\end{figure}

\begin{figure}[htbp]
\includegraphics[width=0.48\textwidth,clip=]{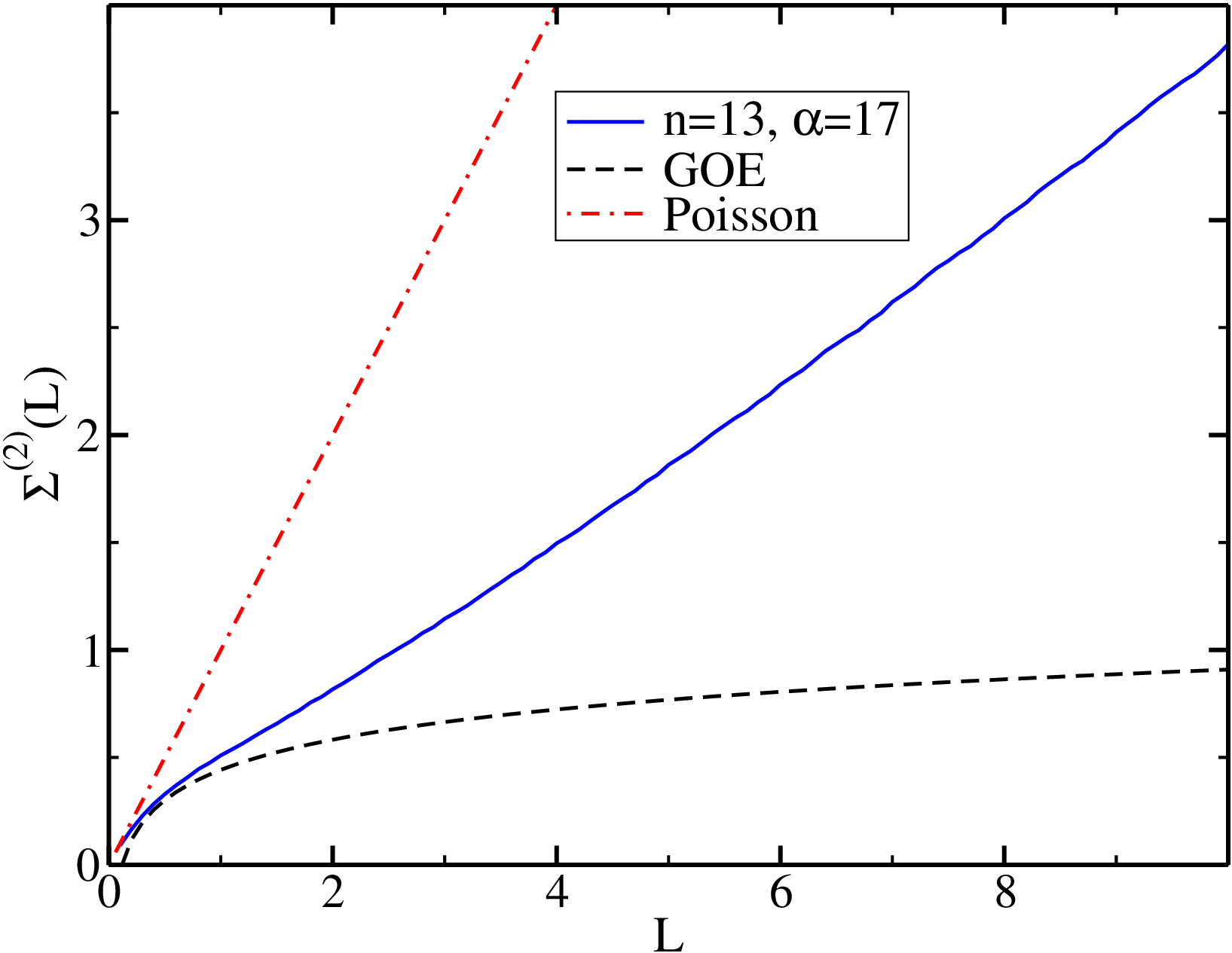}
\caption{(Color online) Long--range $\Sigma^{(2)}$ statistic of the
  scaffolding ensemble versus the length $L$ of the energy interval
  (in units of the mean level spacing) for $n=13$ and $\alpha=17$
  compared to that of a Poissonian spectrum and the GOE.}
\label{fig:sigma2_n13_al17}
\end{figure}

More detailed and extensive numerical calculations show that we
approach the critical regime (the MIT) for $\alpha \approx 17$ and $n
\ge 13$. (Within the available computational resources, these
parameter values optimally satisfy the conditions $n \gg \sqrt{\alpha}
\gg 1$.) The nearest--neighbor spacing distribution shown in
Fig.~\ref{fig:nns_n13_al17} and the long--range $\Sigma^{(2)}$
statistic shown in Fig.~\ref{fig:sigma2_n13_al17} support this
claim. For these parameter values, the levels of the scaffolding
ensemble are clearly strongly correlated at short distances and only
weakly correlated at longer distances.

\begin{figure}[htbp]
\includegraphics[width=0.48\textwidth,clip=]{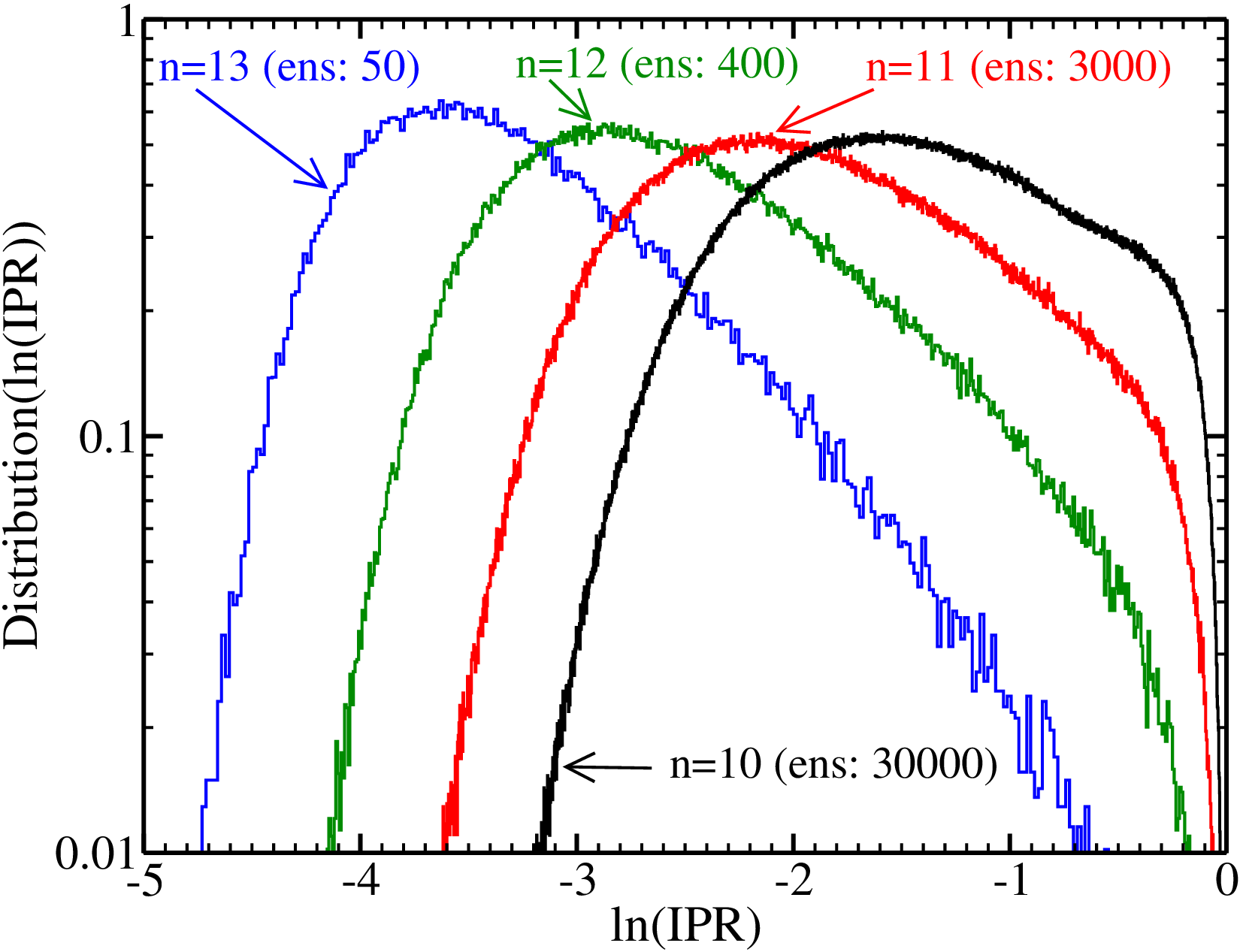}
\caption{(Color online) Distribution of inverse participation ratios
  (IPR) for $\alpha=17$ and $n=10, 11, 12, 13$. As $n$ is increased,
  the distribution approaches a scale--invariant form. The figures
  behind the symbol ``ens'' give the number of realizations.}
\label{fig:ipr_al17}
\end{figure}

The hallmark of the MIT, however, is a scale--invariant distribution
of the inverse participation ratio (IPR)~\cite{Mir00}. The IPR of a
normalized state with expansion coefficients $\psi_\nu$ is defined as
\be
\label{ipr}
{\rm IPR}=\sum_\nu |\psi_\nu|^4 \ .
\ee
Figure~\ref{fig:ipr_al17} shows the distribution of the IPR for the
eigenstates of the scaffolding ensemble. At fixed $\alpha = 17$ and
with increasing $n$, the distribution becomes scale invariant (the
form of the distribution becomes independent of the dimension $N =
2^n$ of the ensemble).  Following Ref.~\cite{Mir00}, we determine the
fractal dimension $D_2$ from the shift of the IPR distribution that
results from a doubling of the dimension as $D_2 \approx 0.85$. For
the largest $n$ shown, the IPR distribution exhibits a power--law tail
with exponent $x_2 \approx 1.5$.

Let us summarize the main results of this Section. We have shown
analytically that the average spectrum of ScE has the shape of a
semicircle. For the three fluctuation measures that characterize the
MIT, our numerical results indicate that ScE possesses critical
statistic for $\alpha \gg 1$ and $n \to \infty$.  This conclusion is
on somewhat safer grounds than are statements for EGOE($k$) simply
because ScE is much more sparse: For $n = 13$, the fraction of
non--zero matrix elements in every row and column is less than $ 2
\times 10^{-3}$ while for EGOE($k$) with $k = 2$ and matrices of
similar dimension, it is about two orders of magnitude bigger. For
the ScE and $n \gg \alpha^{1/2} \gg1$, Levitov's criterion correctly
indicates critical statistic. We conclude that Levitov's criterion
applies to sparse random matrices. We now use that fact to establish
the spectral fluctuation properties of EGOE($k$) for $k > 1$.

\section{Spectral Properties of EGOE($k$)}
\label{spec}

We compare the EGOE($k$) defined in Section~\ref{EGOE} with ScE as
defined in Section~\ref{ScE}. We observe that EGOE($k$) differs from
ScE in three important ways. (i) Counting shows that for all $k = 1,
\ldots, m$ the number of non--vanishing non--diagonal matrix elements
$H_{\mu \nu}$, equal in every row and every column, is given by
$\sum_{p = 1}^k {m \choose p} {l - m \choose p}$ and, thus, for $l,m
\gg 1$ much larger than $\ln N \approx m \ln l$. Hence, for all $k$
EGOE($k$) is much less sparse than ScE. Therefore, we expect EGOE($k$)
to lie on the delocalized side of the MIT and to possess Wigner--Dyson
spectral statistics. (ii) The number of $k$--body matrix elements
$v^{j_1 \ldots j_k}_{j'_1 \ldots j'_k}$ contributing to a fixed
$m$--body matrix element $H_{\mu \nu}$ is, in general, bigger than
one.  For $k = 2$, for instance, the $m$--body matrix element of two
Slater determinants differing in the occupation numbers of orbitals
$1$ and $2$ but both with occupied orbitals $3, 4, \ldots, (m + 1)$
equals $\sum_{j = 3}^{m + 1} v^{1 j}_{2 j}$. In general, the matrix
element connecting two Slater determinants that differ in the
occupation numbers of $p$ single--particle states is the sum of ${m-p
\choose k-p}$ $k$--body matrix elements. We conclude that the
variances of the diagonal elements are all equal and given by $2 {m
\choose k}$; those of the non--vanishing non--diagonal elements
connecting two Slater determinants that differ in the occupation
numbers of $p$ single--particle states are ${m-p \choose k-p}$. It
follows that for $N \to \infty$ the variances of the diagonal elements
are much bigger than those of all non--diagonal elements. That
property is assumed by Levitov's criterion. (iii) $m$--body matrix
elements occurring in different rows and columns may be
correlated. For $k = 2$, for instance, the matrix element $v^{1 2}_{3
4}$ contributes to all $m$--body matrix elements of pairs of Slater
determinants for which the occupation numbers of orbitals ($1$ and
$2$) and ($3$ and $4$) differ while all other occupation numbers
agree. Correlations occur only among $m$--body matrix elements in
different rows and columns because the same $k$--body matrix element
cannot connect a given Slater determinant with two different ones.

{\it Neglect of Correlations.} We first consider EGOE($k$) under
neglect of all correlations, i.e., without property (iii), so that
$k$--body matrix elements appearing in different locations of the
$m$--body EGOE matrix are assumed to be uncorrelated Gaussian random
variables.  This assumption greatly increases the number of
independent random variables and destroys the connection between the
resultant random--matrix ensemble and the random $k$--body Hamiltonian
of Eq.~(\ref{p4}). Under this assumption we conclude immediately that
for $N \to \infty$, the spectral properties of EGOE($k$) are for all
$k$ the same as for GOE. The proof proceeds as for ScE. For the
average level density we use the fact that $\sum_\nu (1 - \delta_{\mu
\nu}) \langle |H_{\mu \nu}|^2 \rangle = \sum_{p = 1}^k {m-p \choose
k-p} {m \choose p} {l-m \choose p}$ increases with $N$ much more
strongly than the variances $2 {m \choose k}$ of the diagonal
elements. That property guarantees that the Pastur equation holds.  The
analogue of the ScE relation $\sum_{\rho} B^{(n)}_{\mu \rho} = \alpha
+ n$ for all $\mu$ also holds: The scaffolding matrix of EGOE($k$) has
the same number of non--zero entries in every row and in every
column. It follows as in Section~\ref{ScE} that the average spectrum
has the shape of a semicircle. The radius $2 \lambda^{{\rm EGOE}(k)}$
is given by $(\lambda^{{\rm EGOE}(k)})^2 = 2 {m \choose k} + \sum_{p =
1}^k {m - p \choose k - p} {m \choose p} {l - m \choose p}$. For the
spectral fluctuations we use Levitov's criterion. We avoid infinitely
large variances for $N \to \infty$ by rescaling the energy and all
matrix elements of EGOE($k$) by the factor $(2 {m \choose k})^{-
1/2}$. Then the diagonal elements all have unit variance, and the
variances ${m - p \choose k - p} / [2 {m \choose k}]$ of the
non--diagonal elements all become very small as $N \to \infty$. For
all $k \geq 1$ the critical sum $S = \sum_{p = 1}^k {m-p \choose k -
p}^{1/2} {m \choose p} {l - m \choose p} / (2 {m \choose k})^{1/2}$
diverges more strongly with $N \to \infty$ than $\ln N \approx m \ln
l$, and Levitov's criterion implies that the spectral statistics is of
Wigner--Dyson type. 

{\it Influence of Correlations.} Correlations among $m$--body matrix
elements occurring in different rows and columns influence the average
level density $\rho(E)$ and the spectral correlations in different
ways. For $\rho(E)$, correlations cause deviations from the
semicircular shape. Indeed, the Pastur equation is derived under the
assumption that cross--linked contributions are negligible. That
assumption fails in the presence of correlations, i.e., when $\langle
H_{\mu \nu} H_{\rho \sigma} \rangle \neq 0$ for $\{\mu, \nu\} \neq
\{\sigma, \rho\}$. Such correlations, nonexistent for $k = m$, become
stronger as $k$ decreases, attaining a maximum at $k = 1$. For
$\rho(E)$ correlations cause cross--linked contributions to be as
important as nested ones. Mon and French~\cite{Mon75}, calculating
even moments of EGUE($k$) in a basis of Slater determinants and using
the representation of ${\cal H}^{(k)}$ in Eq.~(\ref{p4}), have shown
that in the dilute limit and for $k \ll m$ such contributions drive
$\rho(E)$ towards a Gaussian. Thus, the shape of the average spectrum
of EGOE($k$) is expected to change from Gaussian form for $k = 1$
(where correlations are strongest) to semicircular shape for $k = m$
(where correlations are absent).

We do not expect correlations between $m$--body matrix elements
located in different rows and different columns of $H_{\mu \nu}$ to
influence the spectral fluctuations of EGOE($k$). Wigner--Dyson
statistics is a robust property of spectra caused by level repulsion.
Such repulsion is caused by individual matrix elements connecting
pairs of close--lying levels and is independent of the presence of
other correlated matrix elements. Strong support for this qualitative
argument comes from the study of EGOE($1$).

\section{A Special Case: EGOE(1)}

The EGOE($1$) is special: The real--symmetric matrix
$v^j_{j'}$ can be diagonalized, and the eigenvalues follow
Poisson statistics. The $m$--body matrix $H_{\mu \nu}$ is then
diagonal, too, each diagonal element being given by a sum of $m$ such
eigenvalues. For $m \gg 1$ such sums are uncorrelated, and the
spectrum is Poissonian. That symmetry of EGOE($1$) is not obvious in
the $m$--body matrix representation. In excluding such a hidden
symmetry for $k \geq 2$ we appeal to the results of numerical
diagonalizations. Although done for matrices of small dimensions, such
calculations should have revealed the existence of a symmetry.

\begin{figure}[htbp]
\includegraphics[width=0.48\textwidth,clip=]{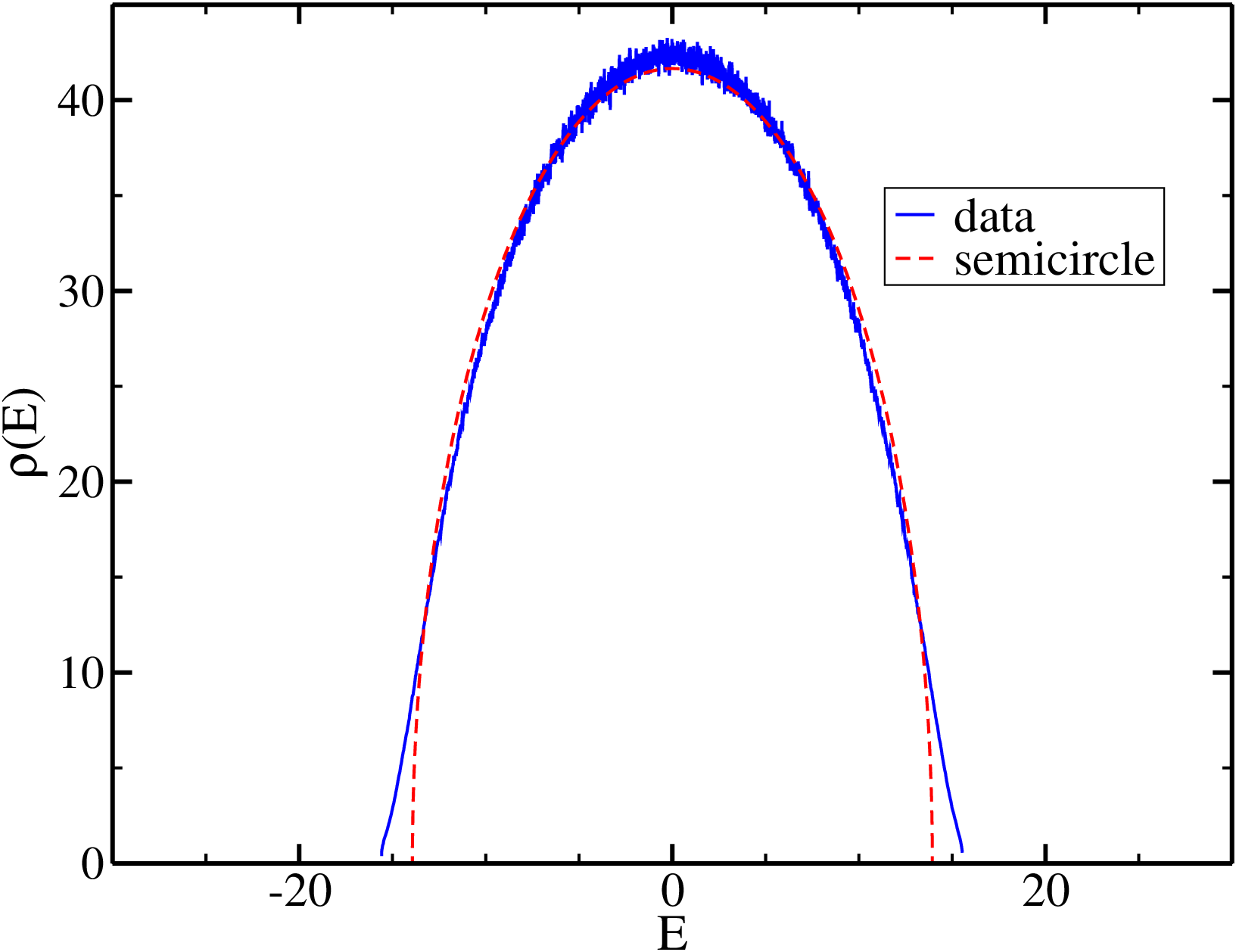}
\caption{(Color online) Level density $\rho(E)$ (compared to a
  best--fit semicircular density) for a random--matrix ensemble with a
  scaffolding matrix of the EGOE(1) and random, uncorrelated matrix
  elements (case (ii) discussed in the text). The ensemble consists of
  1000 realizations for $l = 12$ orbitals and $m = 6$ fermions.}
\label{fig:rho_ii}
\end{figure}

\begin{figure}[htbp]
\includegraphics[width=0.48\textwidth,clip=]{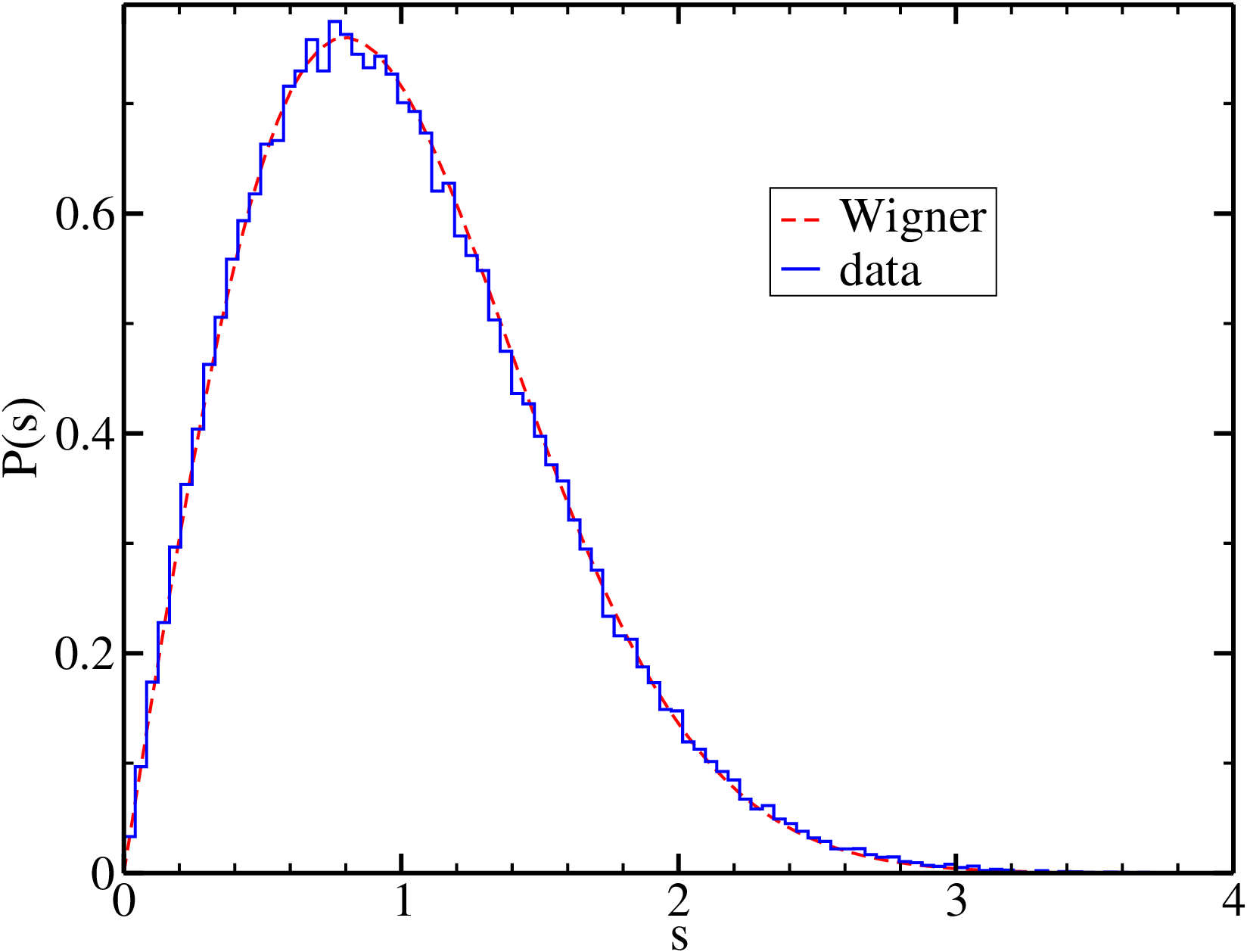}
\caption{(Color online) Nearest--neighbor spacing distribution $P(s)$
  compared to Wigner's surmise for
  a random--matrix ensemble with a scaffolding matrix of the EGOE(1)
  and random, uncorrelated matrix elements (case (ii) discussed in the
  text). The ensemble consists of 1000 realizations for $l = 12$
  orbitals and $m = 6$ fermions.}
\label{fig:nnsd_ii}
\end{figure}

\begin{figure}[htbp]
\includegraphics[width=0.48\textwidth,clip=]{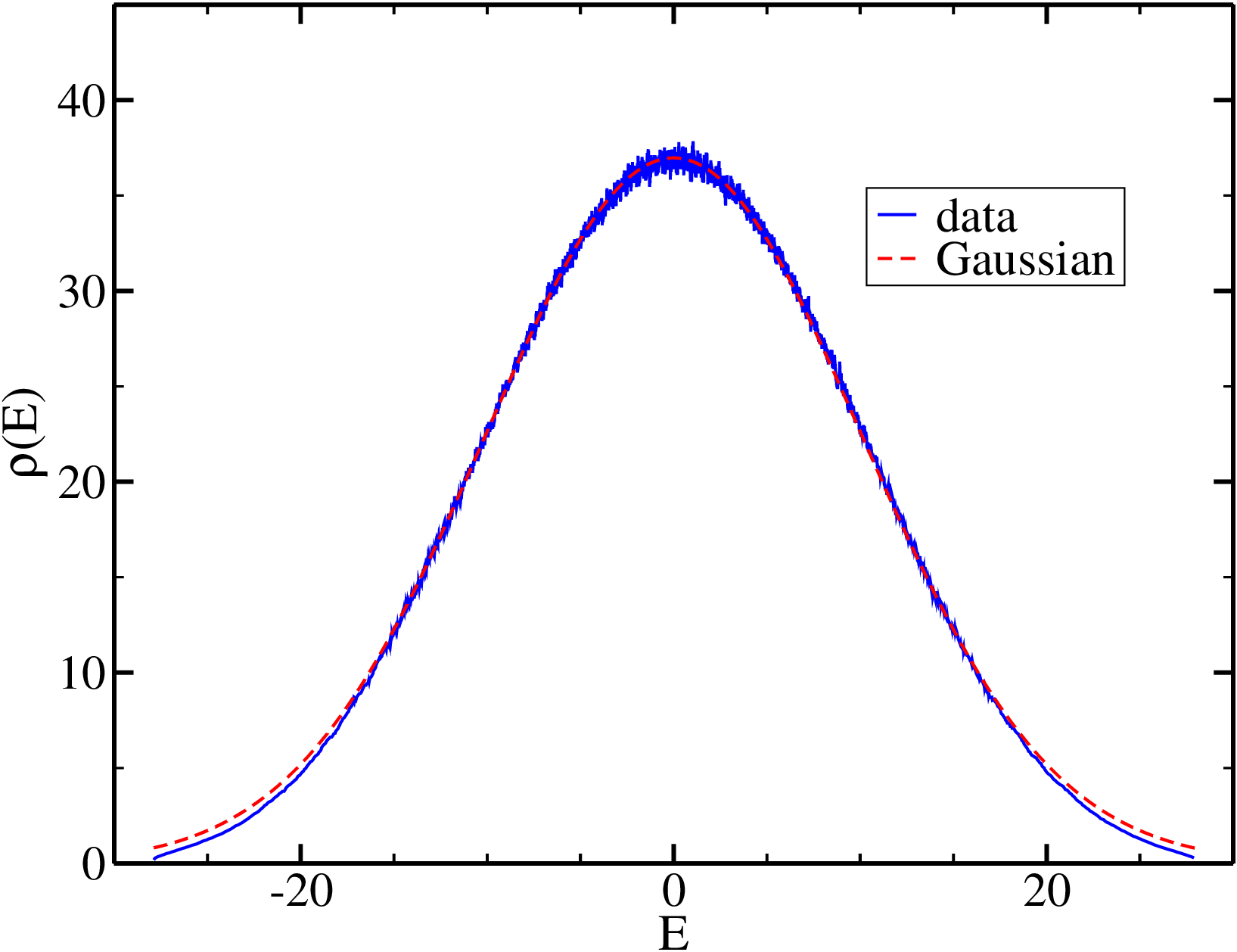}
\caption{(Color online) Average level density $\rho(E)$ (compared to a
  best--fit Gaussian density) for a random--matrix ensemble with a
  scaffolding matrix of the EGOE(1) but with randomized correlated
  matrix elements from the EGOE(1) (case (iii) discussed in the text).
  The ensemble consists of 1000 realizations for $l = 12$ orbitals and
  $m = 6$ fermions.}
\label{fig:rho_iii}
\end{figure}

\begin{figure}[htbp]
\includegraphics[width=0.48\textwidth,clip=]{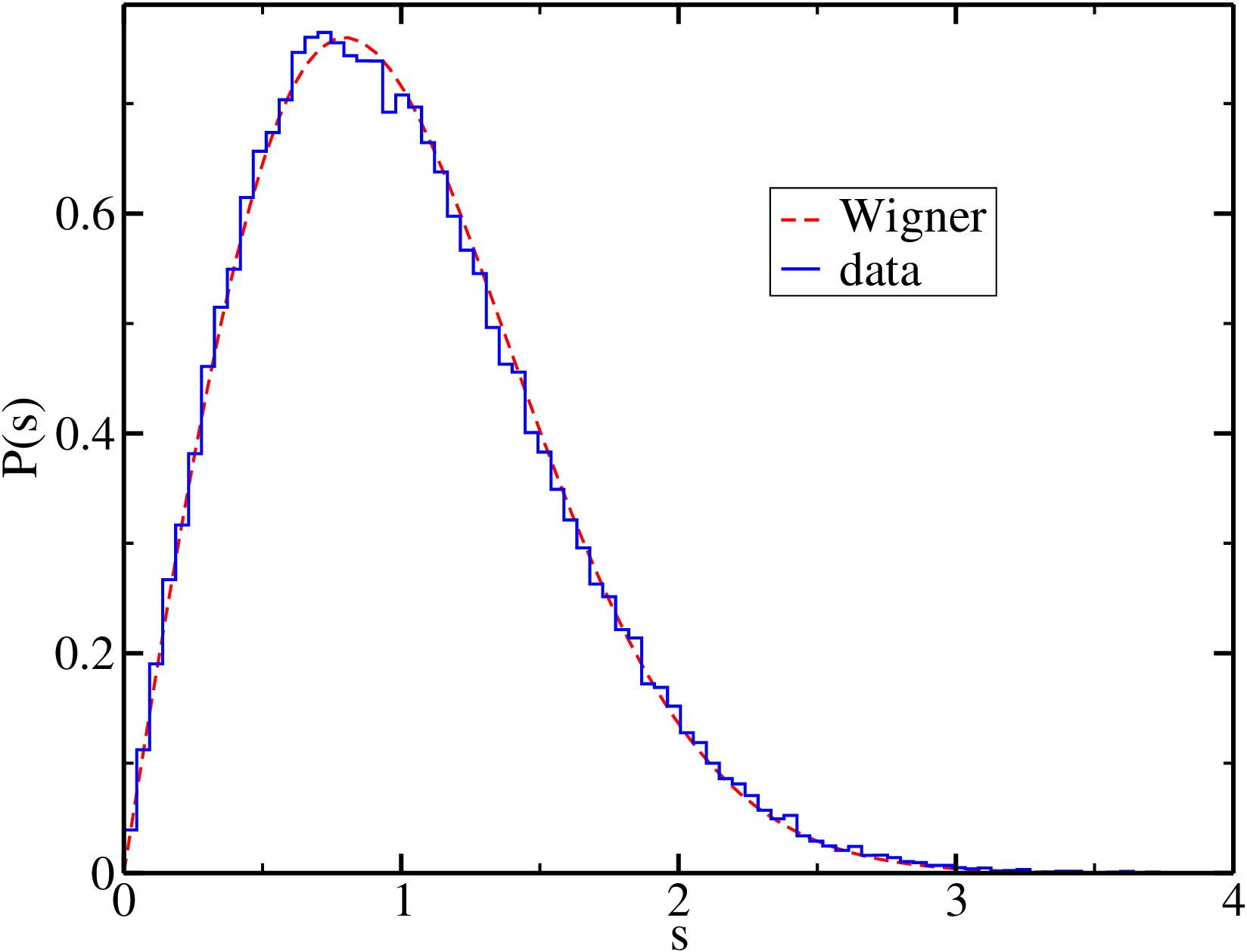}
\caption{(Color online) Nearest--neighbor spacing distribution $P(s)$
  compared to Wigner's surmise for a
  random--matrix ensemble with a scaffolding matrix of the EGOE(1) but
  with randomized correlated matrix elements from the EGOE(1) (case
  (iii) discussed in the text). The ensemble consists of 1000
  realizations for $l = 12$ orbitals and $m = 6$ fermions.}
\label{fig:nnsd_iii}
\end{figure}

Because of that special feature, the case $k = 1$ can be used to
support some of our arguments and conclusions very nicely. We compare
three ensembles. (i) We consider the EGOE($1$). As is well
known~\cite{Mon75,Ben03}, the EGOE($1$) has an average spectrum that
is (nearly) Gaussian, and the eigenvalues have Poisson statistics.
(ii) We consider an ensemble that has the same ``scaffolding matrix''
as EGOE($1$) but for which all independent $m$--body matrix elements
are uncorrelated Gaussian--distributed real random variables. In this
ensemble, the connection with the EGOE($1$) is severed, integrability
is lost, and all correlations present in EGOE($1$) are
destroyed. Because of the lack of correlations between the elements of
the random matrix, we expect the density $\rho(E)$ to have
semicircular shape. Fig.~\ref{fig:rho_ii} shows that this is indeed
the case. Because of the loss of integrability we also expect
Wigner--Dyson statistics for the eigenvalues. This expectation is
confirmed in Fig.~\ref{fig:nnsd_ii}. (iii) A third ensemble is
generated by randomly redistributing the single--particle matrix
elements $v^j_{j'}$ of the Hamiltonian~(\ref{p4}) over the non--zero
elements of the ``scaffolding matrix'' of EGOE($1$). This random
exchange of one--body matrix elements also destroys the connection of
the resulting ensemble with the Hamiltonian~(\ref{p4}) and, thereby,
integrability. However, it retains the existence of correlations
between matrix elements. For ensemble (iii) we, therefore, expect a
(nearly) Gaussian form for the average level density $\rho(E)$ but
Wigner--Dyson statistics for the eigenvalues.  Our numerical
calculations for $l = 12$ and $m = 6$ confirm these expectations:
Figure~\ref{fig:rho_iii} shows that $\rho(E)$ is close to a Gaussian,
and Fig.~\ref{fig:nnsd_iii} confirms the Wigner--Dyson statistics for
the spacing distribution.

We have devoted particular attention to EGOE($1$) not only because it is
special but also because it is much closer to the sparse limit than
EGOE($k$) with $k > 1$. For practicable matrix dimensions it is
difficult to draw valid conclusions about the spectral fluctuation
properties of EGOE($k$) with $k \geq 2$ because the matrices are far
from the sparse limit. Fortunately, correlations are strongest for $k
= 1$. Our conclusions hold, therefore, {\it a fortiori} for $k \geq
2$. In particular, the results for EGOE($1$) strongly support the
conclusions drawn at the end of Section~\ref{spec}.

\section{Conclusions} 

We have constructed a random--matrix ensemble (ScE) that is more
sparse than EGOE($k$) for all $k$. We have shown that ScE mimics the
metal--insulator transition and possesses critical spectral
statistics. Using ScE as a test case, we have verified that Levitov's
criterion applies to sparse random matrices.

Comparison with ScE suggests that for all $k > 1$, EGOE($k$) is on the
delocalized side of the metal--insulator transition and possesses
Wigner--Dyson spectral statistics. We have presented a number of
arguments which strongly support that expectation: Levitov's criterion
indicates chaos, and properties of the (modified) EGOE($1$) illustrate
the different roles played by integrability of the underlying
Hamiltonian on the one hand, and the existence of correlations between
matrix elements on the other. We conclude that spectra in finite
many--body systems governed by few--body interactions generically
display Wigner--Dyson level statistics. By implication we conclude
that in the limit of infinite matrix dimension, the distribution of
the eigenfunctions is Gaussian. These conclusions hold for all three
symmetry classes (orthogonal, unitary, symplectic).

EGOE($k$) is based on the assumption that the single--particle states
are degenerate. If that degeneracy is lifted, chaos may be reduced.
We have not specifically addressed that case~\cite{Jac97}.

We believe that the results of this paper, although not entirely based
on strict analytical arguments, convey new insight into the mechanisms
that determine the spectral shape and the spectral fluctuation
properties of fermionic many--body systems.

ZP thanks J. Kvasil and P. Cejnar for valuable comments. TP and HW
thank O. Bohigas for stimulating discussions. We are grateful to A.
Ossipov for pointing out an error in the original version of the
paper, and for drawing our attention to Ref.~\cite{Lev90}. This work
was supported in parts by the Czech Science Foundation under Grant No.\
202/09/0084 and by the U.S. Department of Energy under Grant No.\
DE-FG02-96ER40963.


\begin{thebibliography}{99}

\bibitem{Guh98}T. Guhr, A. M{\"u}ller-Groeling, and H. A.
Weidenm{\"u}ller, Phys. Rep. {\bf 299} (1998) 189.

\bibitem{Boh84}O. Bohigas, M. J. Giannoni, and C. Schmit, Phys. Rev.
Lett. {\bf 52} (1984) 1.

\bibitem{Heu07}S. Heusler, S. M{\"u}ller, A. Altland, P. Braun, F.
Haake, Phys. Rev. Lett. {\bf 98} (2007) 044103.

\bibitem{Ben03}L. Benet and H. A. Weidenm{\"u}ller, L. Phys. A: Math.
Gen. {\bf 36} (2003) 3569.

\bibitem{Bas08}D. M. Basko, I. L. Aleiner, and B. L. Altshuler, {\it
On the problem of many-body localization}, in: Problems of
Condensed-Matter Physics, A. L. Ivanov and S. G. Tikhodeev, editors,
Oxford University Press, Oxford 2008.

\bibitem{Fyo91}Y. V. Fyodorov and A. D. Mirlin, Phys. Rev. Lett.
{\bf 67} (1991) 2049.

\bibitem{Jac97}Ph. Jacquod and D. L. Shepelyansky, Phys. Rev. Lett.
{\bf 79} (1997) 1837.

\bibitem{And58}P. W. Anderson, Phys. Rev. {\bf 109} (1958) 1492.

\bibitem{Eve08}F. Evers and A. D. Mirlin, Rev. Mod. Phys. {\bf 80}
(2008) 1355.

\bibitem{Mir96}A. D. Mirlin, Y. V. Fyodorov, F.-M. Dittes, J. Quezada,
and T. H. Seligman, Phys. Rev. E {\bf 54} (1996) 3221.

\bibitem{Lev90}L. S. Levitov, Phys. Rev. Lett. {\bf 64} (1990) 547.

\bibitem{Mon75}K. F. Mon and J. B. French, Ann. Phys. (N.Y.) {\bf 95}
(1975) 90.

\bibitem{MS02}
M. Srednicki, Phys. Rev. E {\bf 66}, 046138 (2002). 

\bibitem{Fyo10}Y. V. Fyodorov, A. Ossipov, and A. Rodriguez, J. Stat.
Mech. (2009) L12001.

\bibitem{Mir00}
F. Evers and A. D. Mirlin, Phys. Rev. Lett. 84, 3690 (2000).


\end{thebibliography}
\end{document}